\documentstyle[epsfig]{aipproc}
\begin{document}
\title{Type Ia Supernova Explosion Models:\\
Homogeneity versus Diversity}

\author{Wolfgang Hillebrandt$^*$, Jens C. Niemeyer$^{\dagger}$,
and Martin Reinecke$^*$}
\address{$^*$Max-Planck-Institut f\"ur Astrophysik, Garching, Germany\\
$^{\dagger}$Enrico-Fermi-Institute, University of Chicago, Chicago, IL 60637}

\maketitle

\begin{abstract}
Type Ia supernovae (SN Ia) are generally
believed to be the result of the thermonuclear disruption of  
Chandrasekhar-mass carbon-oxygen white dwarfs, mainly because
such thermonuclear explosions can  account for the right amount
of $^{56}$Ni, which is needed to explain the light
curves and the late-time spectra, and the abundances of intermediate-mass
nuclei which dominate the spectra near maximum light. Because of their
enormous brightness and apparent homogeneity SN Ia  
have become an important tool to measure cosmological 
parameters.
 
In this article the present understanding of the physics of thermonuclear
explosions is reviewed. In particular, we focus our
attention on subsonic (``deflagration'') fronts, i.e. we 
investigate fronts propagating by heat diffusion and convection rather
than by compression. Models based upon this mode of nuclear burning
have been applied very successfully to the SN Ia problem,
and are able to reproduce many of their observed features 
remarkably well. However, the models also indicate that SN Ia
may differ considerably from each other, which is of
importance if they are to be used as standard candles.
\end{abstract}

\section*{Introduction}
Type Ia supernovae, i.e. stellar explosions which do not have hydrogen
in their spectra, but intermediate-mass elements, such as silicon,
calcium, cobalt, and iron, have recently received considerable
attention because it appears that they can be used as ``standard
candles'' to measure cosmic distances out to billions of light years
away from us. Moreover, observations of type Ia supernovae seem to
indicate that we are living in a universe that started to accelerate
its expansion when it was about half its present age. These
conclusions rest primarily on phenomenological models which, however,
lack proper theoretical understanding, mainly because the explosion
process, initiated by thermonuclear fusion of carbon and oxygen into
heavier elements, are difficult to simulate even on supercomputers, 
for reasons we shall discuss in this article.
 
The most popular progenitor model for the average type Ia supernovae 
is a massive white dwarf, consisting of carbon and oxygen,
which approaches the Chandrasekhar mass, M$_{\text{Ch}}$
$~\simeq~\text{1.39}$~M$_{\odot}$,
by a yet unknown mechanism, presumably accretion from a companion star, and
is disrupted by a thermonuclear explosion (see, e.g., \cite{W90}
for a review). Arguments in favor of this hypothesis include the ability of 
the models to fit the observed spectra and light curves rather well.

However, not only is the evolution of massive white dwarfs 
to explosion very uncertain, leaving room for some diversity 
in the allowed set of initial conditions (such as the temperature 
profile at ignition), but also the physics of thermonuclear burning
in degenerate matter is complex and not well understood. The generally
accepted scenario is that explosive carbon burning is ignited 
either at the center of the star or off-center in a couple of ignition
spots, depending on the details of the previous evolution. 
After ignition, the flame is thought to
propagate through the star as a sub-sonic deflagration wave which may or
may not change into a detonation at low densities (around 10$^7$g/cm$^3$).
Numerical models with parameterized velocity of the burning front
have been very successful, the prototype being the W7 model of
\cite{NTY84}. However, these models do
not solve the problem because attempts to determine the effective
flame velocity from direct numerical simulations failed and gave
velocities far too low for successful explosions \cite{L93,K95,AL94a}.
This has led to some speculations about ways
to change the deflagration into a supersonic detonation \cite{K91a,K91b}.
All these aspects of supernova models will be discussed in the 
following sections.

\section*{Nuclear burning in degenerate C+O matter}
Owing to the strong temperature dependence of the nuclear reaction
rates nuclear burning during the 
explosion is  confined to microscopically thin layers that propagate
either conductively as subsonic deflagrations (``flames'') or by shock
compression as supersonic detonations
\cite{CF48,LL95}. Both modes are hydrodynamically unstable
to spatial perturbations as can be shown by linear perturbation
analysis.  In the nonlinear regime, the burning fronts are either
stabilized by forming a cellular structure or become fully turbulent
-- either way, the total burning rate increases as a result of flame
surface growth \cite{LE61,W85,ZBLe85}. Neither flames nor detonations
can be resolved in explosion simulations on stellar scales and
therefore have to be represented by numerical models.

When the fuel exceeds a critical temperature $T_{\rm c}$ where burning
proceeds nearly instantaneously compared with the fluid motions 
a thin reaction zone forms at the interface between burned and unburned
material. It propagates into the surrounding fuel by one of two
mechanisms allowed by the Rankine-Hugoniot jump conditions: a
deflagration (``flame'') or a detonation.

If the overpressure created by the heat of the burning products is
sufficiently high, a hydrodynamical shock wave forms that ignites the
fuel by compressional heating. Such a self-sustaining combustion front that
propagates by shock-heating is called a detonation. 
If, on the other hand, the initial overpressure is too weak, the
temperature gradient at the fuel-ashes interface steepens until an
equilibrium between heat diffusion and energy generation is reached.  The
resulting combustion front consists of a diffusion zone that heats up
the fuel to $T_{\rm c}$, followed by a thin reaction layer where the
fuel is consumed and energy is generated. It is called a deflagration
or simply a flame and moves subsonically with respect to the unburned
material \cite{LL95}.  Flames, unlike detonations, may therefore be
strongly affected by turbulent velocity fluctuations of the fuel. Only
if the unburned material is at rest, a unique laminar flame speed
$S_{\rm l}$ can be found which depends on the detailed interaction of
burning and diffusion within the flame region, e.g.
\cite{ZBLe85}. Following \cite{LL95}, it can be
estimated by assuming that in order for burning and diffusion to be in
equilibrium, the respective time scales, $\tau_{\rm b} \sim
\epsilon/\dot w$ and $\tau_{\rm d} \sim \delta^2/\kappa$, where
$\delta$ is the flame thickness and $\kappa$ is the thermal
diffusivity, must be similar: $\tau_{\rm b} \sim \tau_{\rm
d}$. Defining $S_{\rm l} = \delta/\tau_{\rm b}$, one finds $S_{\rm l}
\sim (\kappa \dot w/\epsilon)^{1/2}$, where $\dot w$ should be
evaluated at $T \approx T_{\rm c}$ \cite{TW92}.  This is only a crude
estimate due to the strong $T$-dependence of $\dot w$. Numerical
solutions of the full equations of hydrodynamics including nuclear
energy generation and heat diffusion are needed to obtain more
accurate values for $S_{\rm l}$ as a function of $\rho$ and fuel
composition.  Laminar thermonuclear carbon and oxygen flames at high
to intermediate densities were investigated by
\cite{BCM80,IIC82,WW86b}, and, using a variety of different
techniques and nuclear networks, by \cite{TW92}. For the purpose of
SN Ia explosion modeling, one needs to know the laminar flame speed
$S_{\rm l} \approx 10^7 \dots 10^4$ cm s$^{-1}$ for $\rho \approx 10^9
\dots 10^7$ g cm$^{-3}$, the flame thickness $\delta = 10^{-4} \dots
1$ cm (defined here as the width of the thermal pre-heating layer
ahead of the much thinner reaction front), and the density contrast
between burned and unburned material $\mu = \Delta \rho/\rho = 0.2
\dots 0.5$ (all values quoted here assume a composition of $X_{\rm C}
= X_{\rm O} = 0.5$ \cite{TW92}). The thermal expansion  parameter
$\mu$ reflects the partial lifting of electron degeneracy in the
burning products, and is much lower than the typical value found in
chemical, ideal gas systems \cite{W85}.

\section*{Hydrodynamic instabilities and turbulence}
The best studied and probably most important hydrodynamical effect for
modeling SN Ia explosions is the Rayleigh-Taylor (RT) instability
resulting from the buoyancy of hot, burned fluid with
respect to the dense, unburned material \cite{MA82,MA86,L93,K94,K95,NH95a},
and after more than five decades of
experimental and numerical work, the basic phenomenology of nonlinear
RT mixing is fairly well understood \cite{F51,L55,S84,R84,Y84}:
Subject to the RT instability, small surface perturbations grow until
they form bubbles (or ``mushrooms'') that begin to float upward while
spikes of dense fluid fall down. In the nonlinear regime, bubbles of
various sizes interact and create a foamy RT mixing layer whose
vertical extent $h_{\rm RT}$ grows with time $t$ according to a
self-similar growth law, $h_{\rm RT} = \alpha g (\mu/2) t^2$, where
$\alpha$  is a dimensionless constant ($\alpha \approx 0.05$) and $g$
is the background gravitational acceleration. 

Secondary instabilities related to  the velocity shear along the
bubble surfaces \cite{NH97} quickly lead to the production of  turbulent velocity
fluctuations that cascade from the size of the largest bubbles
($\approx 10^7$ cm) down to the microscopic Kolmogorov scale, $l_{\rm
k} \approx 10^{-4}$ cm where they are dissipated
\cite{NH95a,K95}. Since no computer is capable of resolving this
range of scales, one has to resort to statistical or scaling
approximations of those length scales that are not properly
resolved. The most prominent scaling relation in turbulence 
research is Kolmogorov's law for the cascade of velocity fluctuations,
stating that in the case of isotropy and statistical stationarity, the
mean velocity $v$ of turbulent eddies with size $l$ scales as $v \sim
l^{1/3}$ \cite{K41}. 
Given the velocity of large eddies, e.g. from
computer simulations, one can use this relation to extrapolate the
eddy velocity distribution down to smaller scales under the assumption
of isotropic, fully developed turbulence \cite{NH95a}.
Knowledge of the eddy velocity as a function of length scale is
important to classify the  burning regime of the turbulent combustion
front \cite{NW97,NK97,KOW97}. The ratio of the laminar flame speed
and the turbulent velocity on the scale of the flame thickness, $K =
S_{\rm l}/v(\delta)$, plays an important role: if  $K \gg 1$, the
laminar flame structure is nearly unaffected by turbulent
fluctuations. Turbulence does, however, wrinkle and deform the flame
on  scales $l$ where $S_{\rm l} \ll v(l)$, i.e. above the  {\em Gibson
scale} $l_{\rm g}$ defined by $S_{\rm l} = v(l_{\rm g})$
\cite{P88}. These  wrinkles increase the flame surface area and
therefore the total energy generation rate of the turbulent front
\cite{D40}. In other words, the turbulent  flame speed, $S_{\rm t}$,
defined as the mean overall propagation velocity of the turbulent
flame front, becomes larger than the laminar speed $S_{\rm l}$. If the
turbulence is sufficiently strong,  $v(L) \gg  S_{\rm l}$, the
turbulent flame speed becomes independent of the laminar speed, and
therefore of the microphysics of burning and diffusion, and scales
only with the velocity of the largest turbulent eddy \cite{D40,C94}:
\begin{equation}
\label{st}
S_{\rm t} \sim v(L)\,\,.
\end{equation}
Because of the unperturbed laminar flame properties on very small
scales, and the wrinkling of the flame on large scales, the burning
regime where $K \gg 1$ is called the corrugated flamelet regime
\cite{P90,C94}. 

As the density of the white dwarf material declines and the laminar
flamelets become slower and thicker, it is plausible that at some
point turbulence significantly alters the thermal flame structure
\cite{KOW97,NW97}. This marks the end of the flamelet regime and
the beginning of the distributed burning, or distributed reaction
zone, regime, e.g. \cite{P90}. So far, modeling the distributed burning
regime in exploding white dwarfs has not been attempted explicitly
since neither nuclear burning and diffusion nor turbulent mixing can
be properly described by simplified prescriptions (see, however,
\cite{LHWe00}). Phenomenologically,
the laminar flame structure is believed to be disrupted by turbulence
and to form a distribution of reaction zones with various lengths and
thicknesses. In order to find the critical density for the transition
between both regimes, we need to formulate a specific criterion for
flamelet breakdown.  A criterion for the transition between both
regimes is discussed in \cite{NW97,NK97} and \cite{KOW97}:
\begin{equation}
l_{\rm cutoff} \le \delta\,\,.
\end{equation} 
Inserting the results of \cite{TW92} for $S_{\rm l}$ and $\delta$ as
functions of density, and using a typical turbulence velocity
$v(10^6\mbox{cm}) \sim 10^7$ cm s$^{-1}$, the transition from flamelet
to distributed burning can be shown to occur at a density of
$\rho_{\rm dis} \approx 10^7$ g cm$^{-3}$ \cite{NK97}.

\section*{Modeling turbulent thermonuclear combustion}
Next we will outline a way by which several of the ideas discussed in
the previous sections can be incorporated into a numerical scheme to
model thermonuclear combustion in SN Ia.

Numerical simulations of any kind of 
turbulent combustion have always been a challenge,
mainly because of the large range of length scales involved. In type
Ia supernovae, in particular, the length scales of relevant
physical processes range from 10$^{-3}$cm for the Kolmogorov-scale
to several 10${^7}$cm for typical convective motions.
In the currently favored scenario the explosion starts as a deflagration
near the center of the star. Rayleigh-Taylor unstable blobs of hot burnt
material are thought to rise and to lead to shear-induced turbulence
at their interface with the unburnt gas. This turbulence increases the
effective surface area of the flamelets and, thereby, the rate of fuel
consumption; the hope is that finally a fast deflagration 
might result, in agreement with phenomenological models of type
Ia explosions \cite{NTY84}.

Despite considerable progress in the field of modeling turbulent
combustion for astrophysical flows (see, e.g., \cite{NH95a}),
the correct numerical representation of the thermonuclear deflagration front
is still a weakness of the simulations. Methods used up to now are
based on for the reactive-diffusive flame model \cite{K93a},
which artificially stretches the burning region over several grid zones 
to ensure an isotropic flame propagation speed. 
However, the soft transition from fuel to ashes stabilizes
the front against hydrodynamical instabilities on small length scales,
which in turn results in an underestimation of the flame surface area and
--~consequently~-- of the total energy generation rate. Moreover,
because nuclear fusion rates depend on temperature nearly
exponentially, one cannot use the zone-averaged values of the
temperature obtained this way to calculate the reaction kinetics.

Therefore we have decided to use a front tracking method to cure
some of these weaknesses. The method is based on the so-called
\emph{level set technique} which was originally  
introduced by Osher and Sethian \cite{OS88}. They used
the zero level set of a $n$-dimensional scalar function to represent
$(n-1)$-dimensional front geometries. Equations for the time evolution of
such a level set which is passively advected by a flow field are given in
\cite{SSO94}. The method has been extended to allow the tracking of
fronts propagating normal to themselves, e.g. deflagrations and detonations
\cite{SMK97,RHNe99,RHN99}. In contrast to the
artificial broadening of the flame in the reaction-diffusion-approach, this
algorithm is able to treat the front as an exact hydrodynamical discontinuity.
We will demonstrate that such a method can be applied to the supernova
problem  and, in addition, we will show that 
even if one attempts to model the physics of thermonuclear burning on 
unresolved length scales well by physically motivated ``Large Eddy
Simulations'' (LES), one still has to perform calculations with very high
spatial resolution in order to be at least near to a converged solution.
The numerical results presented here were obtained by performing 2D
simulations, but there is no problem extending them to
3D. Such simulations are under way.
 
\section*{Application to the supernova problem}
We have carried out numerical simulations in cylindrical rather 
than in spherical coordinates, mainly because it is much simpler to
implement the level set on a Cartesian (r,z) grid. Moreover, the 
CFL condition is somewhat relaxed in comparison to spherical coordinates.
The grid we used in most of our simulations 
maps the white dwarf onto 256$\times$256 mesh points,
equally spaced for the innermost 226$\times$226 zones by
$\Delta$=1.5$\cdot$10$^6$cm, but increasing by 10\% from zone to zone in
the outer parts. The white dwarf, constructed in hydrostatic
equilibrium for a realistic equation of state, has a central density
of 2.9$\cdot$10$^9$g/cm$^3$, a radius of 1.5$\cdot$10$^8$cm, and a
mass of 2.8$\cdot$10$^{33}$g, identical to the one used by
\cite{NH95a}. The initial mass fractions of C and O
are chosen to be equal, and the total binding energy turns out to be
5.4$\cdot$10$^{50}$erg. At low densities ($\rho \leq 10^7$g/cm$^3$), 
the burning velocity of the front is set
equal to zero because the flame enters the distributed regime and our
physical model is no longer valid. However,
since in reality some matter may still burn 
the energy release obtained in the simulations is
probably somewhat too low.
 
\begin{figure*}
\begin{tabular}{cc}
{\epsfig{file=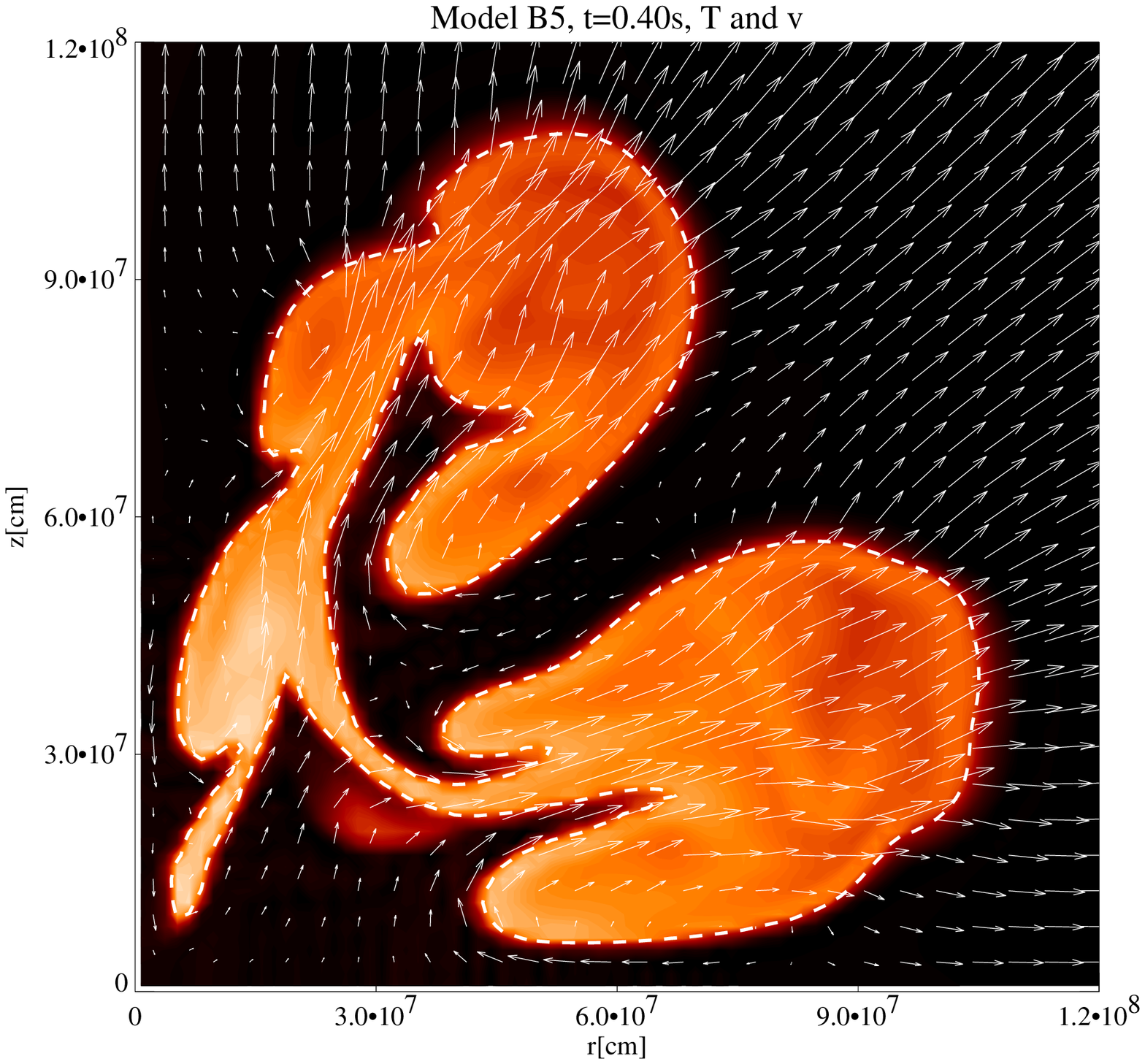, height=0.25\textheight}}
&
{\epsfig{file=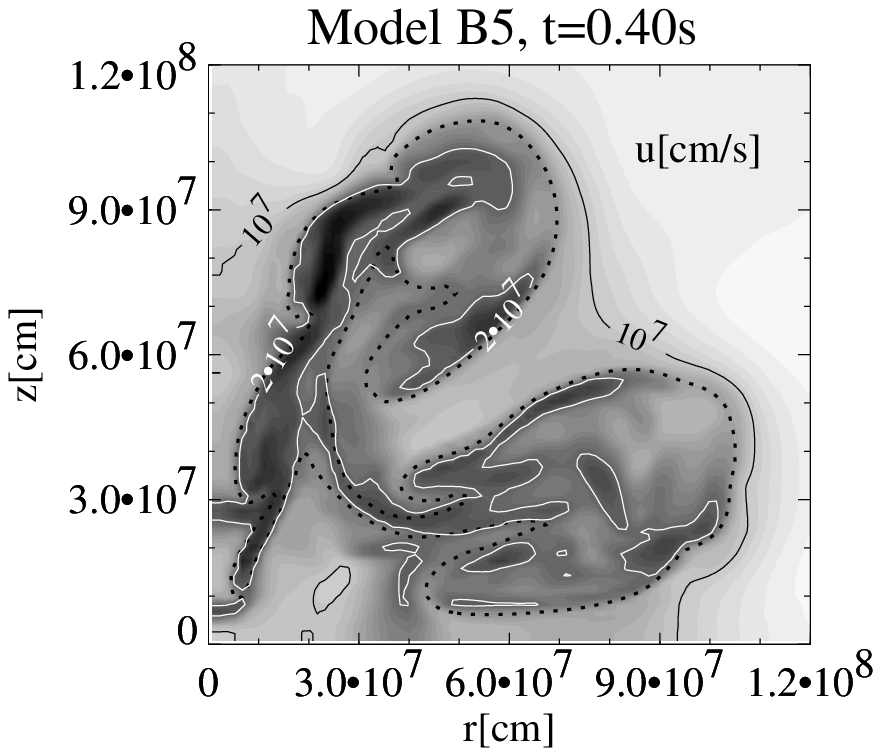, height=0.25\textheight}}
\end{tabular}
\caption{Snapshots of the temperature and the front geometry at 0.4s
for model B5 of Reinecke et al. (1999a) (right figure) and turbulent
velocity fluctuations on the grid scale (left panel). The position of
of the front is indicated by the dotted curve.}
\label{turb}
\end{figure*}

\begin{figure*}
\begin{tabular}{cc}
{\epsfig{file=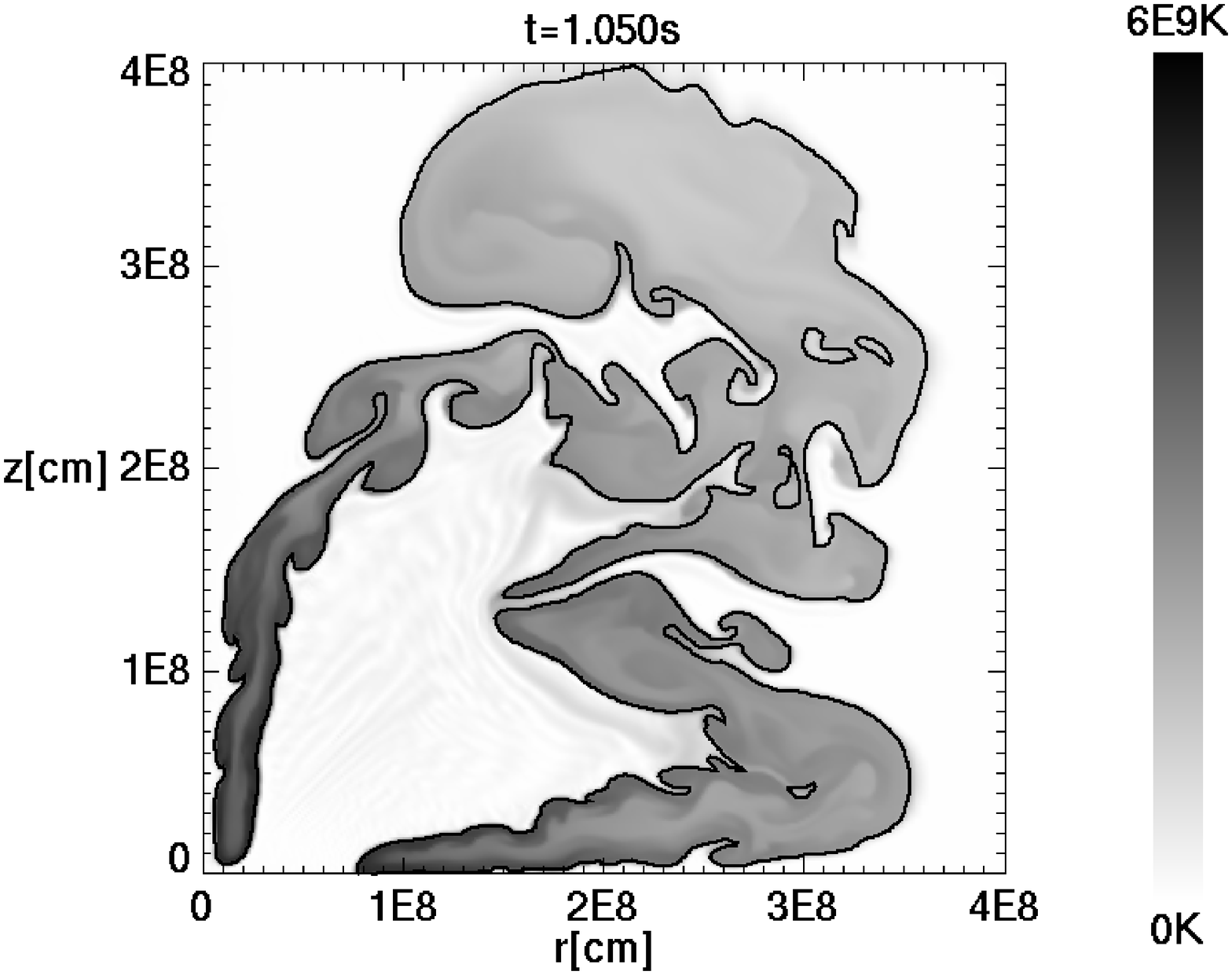, height=0.22\textheight}}
&
{\epsfig{file=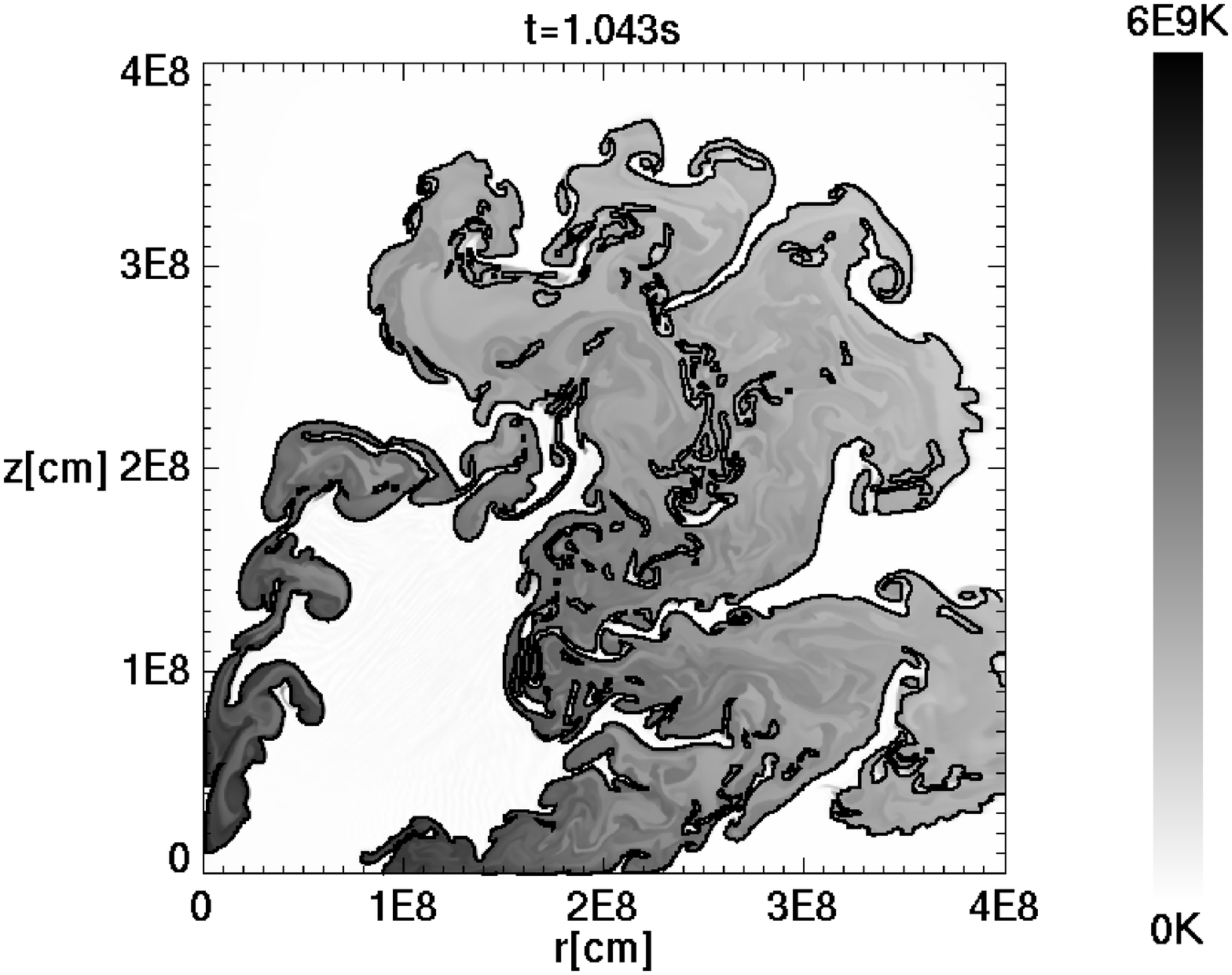, height=0.22\textheight}}
\end{tabular}
\caption{Snapshots of the temperature and the front geometry at 1.05s,
taken for the low-resolution model (left figure) and high-resolution
run, respectively.}
\label{snapsh}
\end{figure*}

Here we present only the results for two models, one in which nuclear burning was
ignited off-center in a blob and a second one in which initially five blobs
were burned as an initial condition, and refer 
to \cite{RHN99} for simulations with other initial conditions.
 
First, in Fig.\ref{turb} a snapshot of the "five-blob" model B5 is shown
at t = 0.4s. The left panel gives the position of the burning front,
the temperatures (in gray-shading), and the expansion velocities.
The right panel shows the distribution of turbulent velocity fluctuations.
We find that, in accord with one's intuition, most of the turbulence is
generated in a very thin layer near the front. Since in the limit
of high turbulence intensity the nuclear flames propagate with the
turbulent velocity it is obvious that this propagation velocity
exceeds the laminar flame speed. However, for most of the initial 
conditions we have investigated this increase was not sufficient to
unbind the star. In fact, model B5 of \cite{RHN99} was the only one
of the set that did explode.
 
Moreover, we show the results obtained with 3 times higher resolution
for comparison. Fig.\ref{snapsh} gives a snapshot taken at 1.05s after
ignition for the low-resolution run (left figure) and the
high-resolution run, respectively. Although the increase in spatial
resolution is only a factor of 3, the right panel shows clearly more
structure. This is an important effect because in the flamelet regime 
the rate of fuel consumption, in first order, 
increases proportional to surface area of the burning front.
The net effect is that the low-resolution model stays bound at the
end of the computations, whereas the better resolved model explodes
with an explosion energy of about 2$\times$10$^{50}$erg. 
Fig.\ref{snapsh} also demonstrates  that the level-set prescription
allows to resolve the structure of the burning front down almost
to the grid scale, thus avoiding artificial smearing of the front
which is an inherent problem of front-capturing schemes. We want to
stress that this gain of accuracy is not obtained at the expense of smaller
CFL time steps because in our hybrid scheme the hydrodynamics is still
done with cell-averaged quantities. 

We also want to stress again, however, that the numerical simulations 
described here, although leaving behind unbound white dwarfs in several 
cases, would not fit the observed spectra of SN Ia, mainly because the
expansion velocity of the partially burnt gas is too low. Whether
this problem is cured once the model calculations are extended to three
spatial dimensions has to be seen, but one can expect significant
changes because of the different ratios of volume to surface area
of burning blobs in 3D.   

\section*{Delayed detonations?}
Turbulent deflagrations can sometimes be observed to undergo
spontaneous transitions to detonations (deflagration-detonation
transitions, DDTs) in terrestrial combustion experiments
\cite{W85}. Thus inspired, it   was suggested
that DDTs may occur in the late phase of the explosion, providing
an elegant explanation for the initial slow burning required to
pre-expand the star, followed by a fast combustion mode that produces
large amounts of high-velocity intermediate mass elements
\cite{K91a,WW94a}. Many 1D simulations have meanwhile demonstrated
the capability of the delayed detonation scenario to provide excellent
fits to SN Ia spectra and light curves \cite{W90,HK96}, as well as
reasonable nucleosynthesis products with regard to solar abundances
\cite{K91b,IBNe99}. In the best fit models, the initial flame phase
has a  rather slow velocity of roughly one percent of the sound speed
and transitions to detonation at a density of $\rho_{\rm DDT} \approx
10^7$ g cm$^{-3}$ \cite{HK96,IBNe99}. The transition density was also
found to be a convenient parameter to explain the observed sequence of
explosion strengths \cite{HK96}.

Various mechanisms for DDT were discussed in the early literature
on delayed detonations (see \cite{NW97} and references
therein). Recent investigations have focussed on the induction time
gradient mechanism \cite{ZLMe70,LKY78}, analyzed in the context of
SNe Ia by \cite{BK86} and \cite{BK87}.  It was realized by
\cite{KOW97} and \cite{NW97} that a necessary criterion for this
mechanism is the local disruption of the flame sheet by turbulent
eddies, or, in other words, the transition of the burning regime from
``flamelet'' to ``distributed'' burning. Simple
estimates \cite{NK97} show that this transition should occur at
roughly $10^7$ g cm$^{-3}$, providing a plausible explanation for the
delay of the detonation. 

The critical length (or mass) scale over which the temperature
gradient must be held fixed in order to allow the spontaneous
combustion wave to turn into a detonation was computed by
\cite{KOW97} and \cite{NW97}; it is a few orders of magnitude
thicker than the final  detonation front and depends very sensitively
on composition and density.  Recently, this conclusion was confirmed by
\cite{LHWe00}, who computed the interaction between turbulence and 
the nuclear flame in the distributed regime by means of a particular 
turbulence model, and by \cite{LHW00}.

The virtues of the delayed detonation scenario can be summarized
as follows. It is undisputed
that suitably tuned delayed detonations satisfy all the constraints
given by SN Ia spectra, light curves, and nucleosynthesis. If
$\rho_{\rm DDT}$ is indeed determined by the transition of burning
regimes  -- which in turn might be composition dependent
\cite{UNKe99} --
the scenario is also fairly robust and $\rho_{\rm DDT}$ may represent
the explosion strength parameter. Note that in this case, the
variability induced by multi-point ignition needs to be explained
away. If, on the other hand, thermonuclear flames are  confirmed to be
almost unquenchable, the favorite mechanism for DDTs becomes
questionable \cite{N99}. Moreover, should the mechanism DDT rely on
rare, strong turbulent fluctuations one  must ask about those events
that fail to ignite a detonation following the slow deflagration phase
which, on its own, cannot give rise to a viable SN Ia explosion. They
might end up as pulsational delayed detonations or as unobservably
dim, as yet unclassified explosions. Multidimensional simulations of
the turbulent flame phase  may soon answer whether the turbulent flame
speed is closer to 1 \% or 30 \% of the speed of sound and hence
decide whether DDTs are a necessary ingredient of SN Ia explosion
models.

\section*{Summary and conclusions}
In this review we have outlined our present understanding
of Type Ia supernovae explosions.
From the tremendous amount of work carried out over the last couple
of years it has become obvious that the physics of SNe Ia is very
complex, ranging from the possibility of very different
progenitors to the complexity of the physics leading to the
explosion and the complicated processes which couple the interior
physics to observable quantities. None of these problems is fully
understood yet, but what one is tempted to state is that, from a 
theorist's point of view, it appears to be a miracle that all 
the complexity seems to average out in a mysterious way 
to make the class so homogeneous. In contrast, as it stands,
a save prediction from theory seems to be that SNe Ia should get
more divers with increasing observed sample sizes. 
If, however, homogeneity would continue to hold
this would certainly add support to the Chandrasekhar-mass 
scenario discussed in this article. On the other hand, even an 
increasing diversity would not rule out Chandrasekhar-mass
progenitors for most of them. In contrast,
there are ways to explain how the diversity  
is absorbed in a one parameter family of transformations, such as
the Phillips-relation \cite{P93,HPSe96a} 
or modifications of it \cite{RPK96,PGGe97}. 

As far as the explosion/combustion physics and the numerical
simulations are concerned significant recent progress has made 
the models more realistic (and reliable). Thanks to ever increasing
computer resources 3-dimensional simulations have become feasible
which treat the full star with good spatial resolution and realistic
input physics. Already the results of 2-dimensional simulations
indicate that pure deflagrations waves in Chandrasekhar-mass C+O white 
dwarfs can lead to explosions, and one can expect that going to
three dimensions, because of the increasing surface area of the
nuclear flames, should add to the explosion energy. If confirmed,
this would eliminate pulsational detonations from the list of
potential models. On the side of the combustion physics, the 
burning in the distributed regime at low densities needs to be
explored further, but it is not clear anymore whether a transition
from a deflagration to a detonation in that regime is needed for
successful models. In fact, according to recent studies such a transition 
appears to be rather unlikely.

{\bf Acknowledgments:}
This work was supported in part by the Deutsche Forschungsgemeinschaft under
Grant Hi 534/3-1, the DAAD, and by DOE under contract No. B341495 at the University of
Chicago. The computations were performed at the Rechenzentrum Garching
on a Cray J90.

\bibliography{HNR-contr}
 
\end{document}